\documentclass{IEEEtran}
\usepackage{cite}
\usepackage{amsmath,amssymb,amsfonts}
\usepackage{graphicx}
\usepackage{graphbox}
\usepackage{textcomp,nicefrac}
\usepackage{scalerel}
\usepackage{siunitx}
\DeclareSIUnit{\neqcm}{n_{eq}}
\usepackage{tikz}
\usetikzlibrary{svg.path}

\definecolor{orcidlogocol}{HTML}{A6CE39}
\tikzset{
  orcidlogo/.pic={
    \fill[orcidlogocol] svg{M256,128c0,70.7-57.3,128-128,128C57.3,256,0,198.7,0,128C0,57.3,57.3,0,128,0C198.7,0,256,57.3,256,128z};
    \fill[white] svg{M86.3,186.2H70.9V79.1h15.4v48.4V186.2z}
                 svg{M108.9,79.1h41.6c39.6,0,57,28.3,57,53.6c0,27.5-21.5,53.6-56.8,53.6h-41.8V79.1z M124.3,172.4h24.5c34.9,0,42.9-26.5,42.9-39.7c0-21.5-13.7-39.7-43.7-39.7h-23.7V172.4z}
                 svg{M88.7,56.8c0,5.5-4.5,10.1-10.1,10.1c-5.6,0-10.1-4.6-10.1-10.1c0-5.6,4.5-10.1,10.1-10.1C84.2,46.7,88.7,51.3,88.7,56.8z};
  }
}
\newcommand\orcidicon[1]{\href{https://orcid.org/#1}{\mbox{\scalerel*{
\begin{tikzpicture}[yscale=-1,transform shape]
\pic{orcidlogo};
\end{tikzpicture}
}{|}}}}

\usepackage{hyperref}
\hypersetup{
  colorlinks=false,
  linkbordercolor=white,
 urlbordercolor=white,
pdfborder={0 0 0}
}

\usepackage[switch]{lineno}
\modulolinenumbers[5]

\def\BibTeX{{\rm B\kern-.05em{\sc i\kern-.025em b}\kern-.08em
T\kern-.1667em\lower.7ex\hbox{E}\kern-.125emX}}
\DeclareSIUnit{\neqcm}{\text{n}_\text{eq} \text{/} \text{cm}^\text{2}}

\begin{document}
\title{Cryogenic operation of neutron-irradiated silicon photomultiplier arrays up to \SI{1e14}{\neqcm}}

\author{
    E.~Currás-Rivera$^{\textsuperscript{\orcidicon{0000-0002-6555-0340}}}$,
    G.~Haefeli$^{\textsuperscript{\orcidicon{0000-0002-9257-839X}}}$ and 
    F.~Ronchetti$^{\textsuperscript{\orcidicon{0000-0003-3438-9774}}}$
    
\thanks{E.~Currás-Rivera, G.~Haefeli and F.~Ronchetti are with the Institute of Physics, École Polytechnique Fédérale de Lausanne}}

\maketitle

\begingroup
\renewcommand\thefootnote{}\footnote{This work has been submitted to the IEEE for possible publication. Copyright may be transferred without notice, after which this version may no longer be accessible.}
\addtocounter{footnote}{-1}
\endgroup

\begin{abstract}
In the context of the Scintillating Fibre (SciFi) Tracker for the LHCb Upgrade~2, radiation-induced damage in silicon photomultipliers (SiPMs) has been studied over a wide temperature range, from room temperature down to 100~K.
With the LHCb detector Upgrade~1, installed during the LHC’s Long Shutdown~2 (LS2) (2019–2021), the first large-scale SciFi tracker read out by multichannel silicon photomultipliers (SiPMs) was constructed, installed, and has been operated ever since.
A major challenge for the SciFi tracker is the neutron radiation at the SiPMs’ location. At the end of the lifetime of the Upgrade~1 detector, the expected neutron fluence for the SiPMs will reach \SI{6e11}{\neqcm}. Cryogenic operation is being investigated to mitigate even higher radiation-induced damage for Upgrade~2, where the total neutron fluence is expected to reach \SI{3e12}{\neqcm}.
A large set of custom SiPM arrays, varying in pixel size, electric field configuration, and doping implant concentration, developed by FBK and Hamamatsu were tested after neutron irradiation. Characterisation was performed in a dedicated cryogenic test setup, where the operating temperature was varied over a wide range. Key performance parameters such as breakdown voltage, gain, dark count rate, optical crosstalk, and afterpulsing were characterised as functions of temperature, overvoltage, and neutron fluence. The result is a precise assessment of radiation damage for state-of-the-art technology from two leading SiPM manufacturers, allowing the results to be transferred to other SiPM applications.
\end{abstract}


\section{Introduction}
\label{sec:introduction}

For the LHCb Upgrade~2 at the Long Shutdown~4 (LS4)~\cite{CERN-LHCC-2024-010}, the detector will undergo a major upgrade to increase the instantaneous luminosity and with this, the expected integrated luminosity will reach \SI{300}{fb^{-1}} for the entire High-Luminosity LHC programme. The current Scintillating Fibre (SciFi) tracking detector will be replaced by a hybrid system consisting of a CMOS pixel detector close to the beam pipe, in the highest track density region, and a SciFi tracker in the outer region.  

A major challenge for the existing SciFi tracker is the radiation environment, where SiPMs are exposed to a neutron fluence of up to \SI{3e12}{\neqcm}. The radiation environment is dominated by fast neutrons originating from the hadronic calorimeter. To overcome radiation-induced damage to silicon photomultipliers (SiPMs), cryogenic cooling with liquid nitrogen is being investigated as a mitigation strategy~\cite{currasrivera2025,CURRASRIVERA2026170833}.

Over the past decades, SiPMs have undergone major technological improvements, establishing themselves as a viable alternative to traditional photomultiplier tubes for photon detection at room temperature. More recently, the cryogenic operation of SiPMs has attracted growing interest in high-energy physics applications, particularly for detecting single photons in noble-liquid detectors and Cherenkov-based systems~\cite{COLLAZUOL2011389, Sun_2016, ACERBI2023167683}. These developments have driven substantial improvements in SiPM performance and stability at low temperatures.  

Nevertheless, radiation damage remains a major limitation for SiPMs in applications at collider experiments or space missions. More radiation-tolerant SiPM technologies will help extend the application range and improve the performance. Studies for different application fields and conditions have been published on the subject for various non-ionising radiation conditions to quantify their degradation mechanisms and to guide the design of more robust devices~\cite{ALTAMURA2022167284, ACERBI2023167791, Addesa_2024}.  

In this work, we present a comprehensive characterisation of the noise performance of SiPMs over a temperature range from near room temperature, 250~K, down to 100~K. The focus of the study is on neutron-irradiated SiPMs fabricated using two distinct technologies from two different manufacturers, Fondazione Bruno Kessler (FBK) and Hamamatsu Photonics K.K. (HPK). The devices were irradiated up to \SI{1e14}{\neqcm} and systematically characterised as functions of temperature, overvoltage, and neutron fluence. The results offer detailed insight into the temperature dependence of key performance parameters and evaluate the potential of cryogenic operation as an effective strategy to mitigate radiation-induced degradation in future LHCb detector upgrades.

\section{SiPM arrays under study}
\label{sec:sipm_arrays}

For clarity, in this work we distinguish between \emph{SiPM arrays} and \emph{SiPM modules}. An \emph{SiPM array} refers to the multi-channel silicon sensor (silicon die) itself, while an \emph{SiPM module} denotes a complete unit consisting of two SiPM arrays soldered or bonded to a PCB together with the associated electronics components and mechanical support.

\begin{figure}[t]
\centering
\includegraphics[width=0.48\columnwidth]{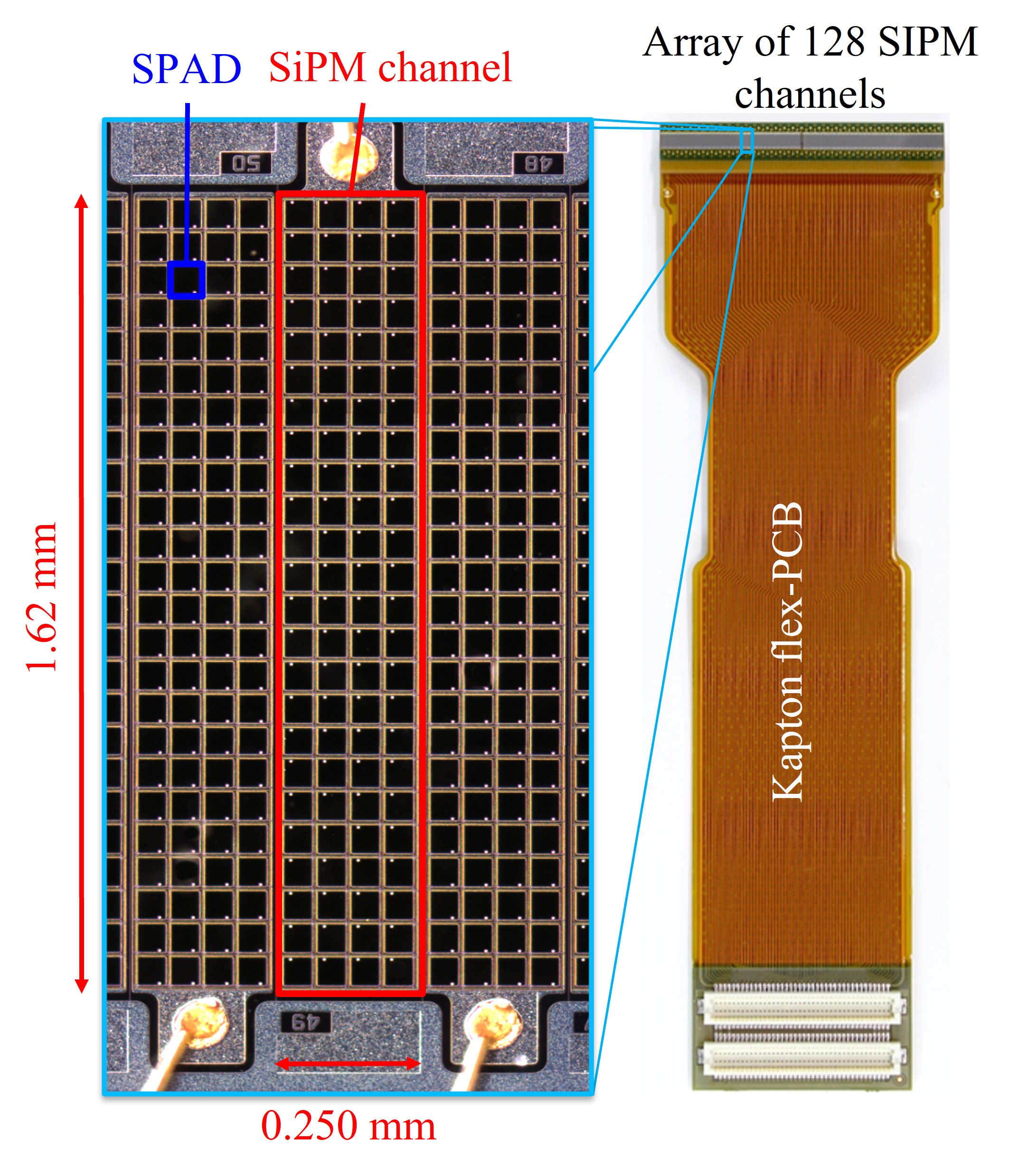}\hfil
\includegraphics[width=0.28\columnwidth]{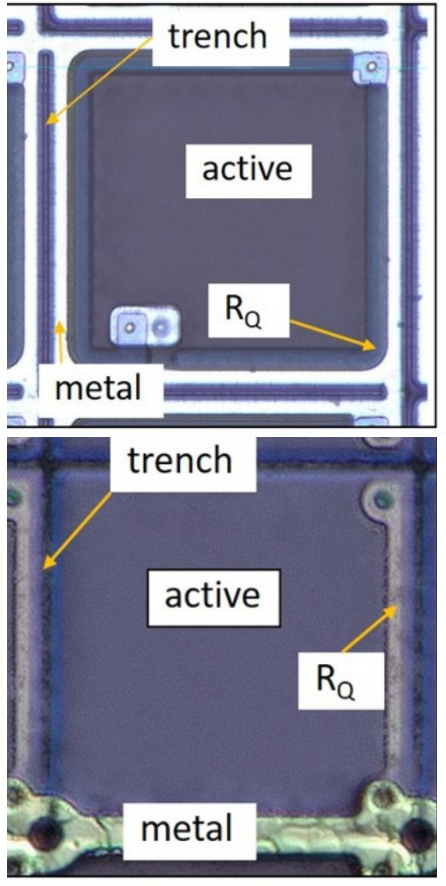}\hfil
\caption{Left: HPK SiPM module installed in the LHCb SciFi tracker during Upgrade~I (produced 2017). Right: layouts of the two SiPM pixel technologies under study, with the HPK design on top and the FBK design on the bottom.} 
\label{SiPM_modules}
\end{figure}

The SiPM arrays were produced by FBK in 2022 and HPK in 2024. They have almost identical active surfaces, but there are small differences that allow for different pixel sizes. The differences are less than 2\% in the total active area, and, for simplicity, we quote the nominal size. Each SiPM array contains 64 channels and features nominal channel dimensions of \SI{0.25}{mm} $\times$ \SI{1.65}{mm}. The detailed values of the pixel size, the channel size, and the total pixel number for each array are given in Tab.~\ref{SiPM_parameters}. An SiPM module contains two silicon chips that are assembled into modules according to two different packaging options, depending on the producer. For FBK arrays, they are directly wire-bonded to the Kapton flex PCB, whereas for HPK arrays, an interposer PCB is used (delivered by HPK) and a reflow soldering process is used to assemble the modules. A temperature sensor (Pt1000) is integrated into each module for local temperature monitoring close to the silicon.  A picture of an HPK SiPM module and an SiPM channel is shown in Fig.~\ref{SiPM_modules} (left). 

 A quenching resistor ($R_q$) of approximately \SI{500}{\kilo\ohm} at room temperature is implemented in all arrays, leading to a recovery time depending on both pixel size (pixel capacitance $C_{pix}$) and temperature ($\tau_{rec}=R_q (T) \cdot C_{pix}$). The quenching resistor material is polysilicon for FBK devices and a thin metal layer for HPK devices. FBK SiPMs are based on the recent near-ultraviolet high-density metal-in-trench (NUV-HD-MT) technology~\cite{Acerbi2017}, specifically from the ultra-low-field (ULF) production optimised for cryogenic operations (produced in 2022). HPK SiPMs carry the production code ID S17104, but no specific technology details are available (produced in 2024). These devices feature through-silicon via (TSV) technology for packaging. Two microscope pictures of the typical pixel layouts are shown in Fig.~\ref{SiPM_modules} (right).
\begin{table}[t]
\begin{center}
 \begin{tabular}{| l | r | r | r | r |} 
 \hline
 \multicolumn{1}{|c|}{SiPM type} & \multicolumn{1}{c|}{Pixel size} & \multicolumn{1}{c|}{Channel size} & \multicolumn{1}{c|}{\#Pix/ch} & \multicolumn{1}{c|}{G/$\Delta$V} \\ 
 \multicolumn{1}{|c|}{} & \multicolumn{1}{c|}{[$\mu$m$^2$]} & \multicolumn{1}{c|}{[$\text{mm}^2$]} & \multicolumn{1}{c|}{} & \multicolumn{1}{c|}{[$\times10^5$]} \\ [0.7ex]
 \hline\hline
 HPK-60 (2017) & $57.5\times62.5$  & $1.625\times0.25$ & 104 & 1.0 \\[0.5ex] 
 \hline
 HPK-50 (2024) & $50.0\times50.0$  & $1.625\times0.25$ & 160 & 7.5 \\[0.5ex] 
 \hline
 HPK-42 (2024) & $41.67\times41.67$  & $1.625\times0.25$ & 234 & 4.9 \\[0.5ex] 
 \hline
 FBK-42 (2022) & $41.73\times41.73$ & $1.627\times0.2504$ & 234 & 8.5 \\[0.5ex] 
 \hline
 FBK-31 (2022) & $31.3\times31.3$  & $1.659\times0.2504$ & 424 & 5.1 \\[0.5ex] 
 \hline
\end{tabular}
\end{center}
\caption{Summary of the parameters for the SiPM arrays used in this study. The gain values correspond to the slope of the charge versus overvoltage curve measured at room temperature.} 
\label{SiPM_parameters}
\end{table}

Neutron irradiation was performed at the Jožef Stefan Institute (JSI) in Ljubljana (Slovenia) at four fluences: \SI{3e11}{\neqcm}, \SI{1e12}{\neqcm}, \SI{3e12}{\neqcm}, and \SI{1e13}{\neqcm}. Additionally, for the HPK arrays, irradiation was extended to \SI{3e13}{\neqcm} and \SI{1e14}{\neqcm}. These fluences were achieved with an uncertainty of 10\,\%. All SiPMs were annealed after irradiation for two weeks at $30\,^{\circ}\mathrm{C}$, therefore the results presented here refer to post-annealing performance. Based on measurements from LHCb Upgrade 1, this annealing is expected to reduce the dark count rate (DCR) by a factor of about 3.3. 

\section{Setup and measurement conditions}
\label{sec:setup}

The setup for the measurements is a closed-cycle helium cryostat with a temperature controller. The cryostat chamber has enough space to accommodate up to four SiPM arrays simultaneously, allowing measurements over a temperature range from 100~K to 300~K (room temperature, RT). The cold head is in contact with a copper support that has two purposes: to provide good thermal contact with the SiPM modules and to allow precise temperature control via two \mbox{50~W} heaters placed underneath. 

The vacuum inside the cryostat chamber is created using a pump system, and the vacuum level and temperature on the cold head, copper support, and SiPM modules under study are always monitored. Several feed-throughs allow access to the chamber to readout the SiPMs under test, to monitor the temperature, and to allow light injection.

For light injection, a laser with a fixed wavelength of \mbox{450 nm} is used, and the light is diffused to ensure uniform illumination. Light injection is employed for the breakdown voltage measurement. For biasing, a Keithley 2450 source meter is used to apply the voltage and monitor the current. A 128-channel multiplexer allows selection between different SiPM channels. The signal is amplified using a 40~dB, 2~GHz bandwidth amplifier and digitized with a 4~GHz, 20~GS/s oscilloscope for recording.

\section{Results}
\label{sec:results}

\subsection{Breakdown voltage}
\label{sec:breakdown}

The breakdown voltage (\(V_{\mathrm{bd}}\)) was extracted from the current--voltage (I--V) characteristics using the Inverse Logarithmic Derivative (ILD) method~\cite{CHMILL201756}. For each temperature point, I--V curves were acquired with laser illumination to improve measurement precision and signal-to-noise ratio. Multiple channels of each SiPM module were measured simultaneously to increase the effective photodetector area under test and reduce fluctuations. As a consequence, the result for \(V_{\mathrm{bd}}\) corresponds to an average value over typically 10 neighbouring channels on the same silicon array. The difference of \(V_{\mathrm{bd}}\) between channels is found to be very small, well below the measurement uncertainty, and therefore confirms the excellent uniformity of the SiPM implementation.

\begin{figure}[t]
\centering
\includegraphics[width=0.80\columnwidth]{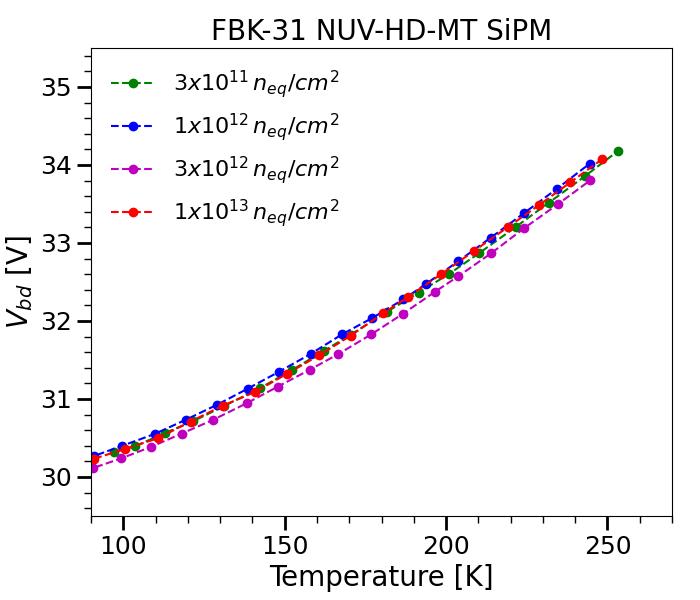}
\caption{Breakdown voltage as a function of temperature for the FBK SiPM module with 31~µm pixel size.}
\label{fig:vbd_FBK31}
\end{figure}

\begin{figure}[t]
\centering
\includegraphics[width=0.80\columnwidth]{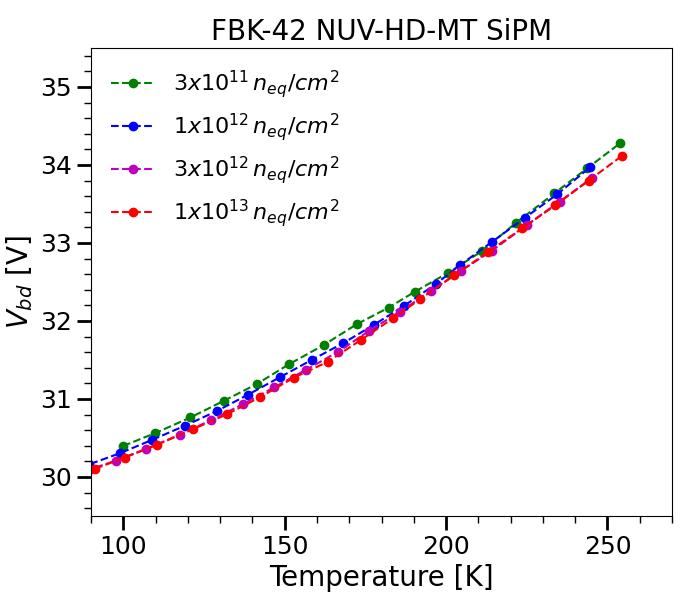}
\caption{Breakdown voltage as a function of temperature for the FBK SiPM module with 42~µm pixel size.}
\label{fig:vbd_FBK42}
\end{figure}

\begin{figure}[t]
\centering
\includegraphics[width=0.80\columnwidth]{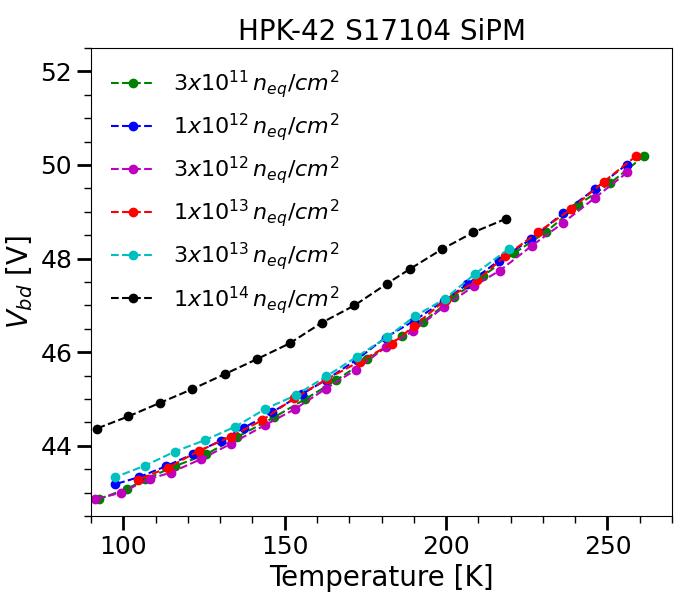}
\caption{Breakdown voltage as a function of temperature for the HPK SiPM module with 42~µm pixel size.}
\label{fig:vbd_HPK42}
\end{figure}

\begin{figure}[t]
\centering
\includegraphics[width=0.80\columnwidth]{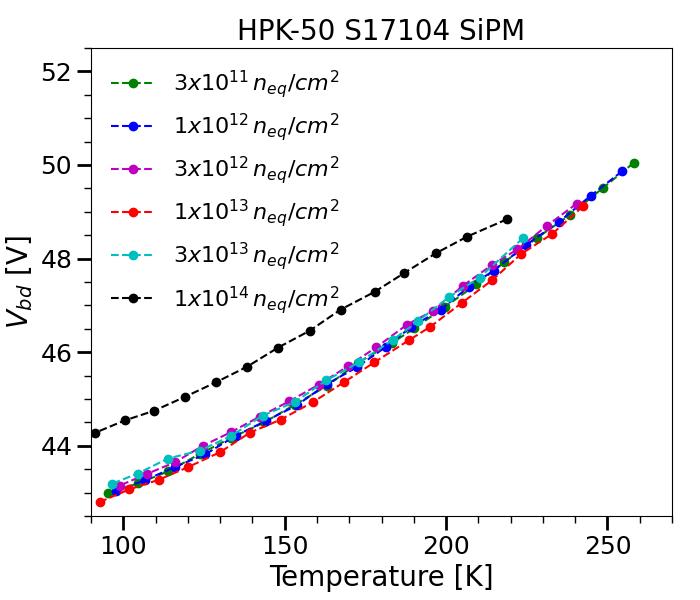}
\caption{Breakdown voltage as a function of temperature for the HPK SiPM module with 50~µm pixel size.}
\label{fig:vbd_HPK50}
\end{figure}

The temperature dependence of \(V_{\mathrm{bd}}\) for the four types of SiPM is shown in Fig.~\ref{fig:vbd_FBK31}--\ref{fig:vbd_HPK50}. In addition to intrinsic device-to-device variation within modules of the same production batch, no significant change in \(V_{\mathrm{bd}}\) is observed after neutron irradiation up to a fluence of \SI{1e13}{\neqcm}. For the highest fluences studied, \SI{3e13}{\neqcm} and \SI{1e14}{\neqcm}, an increase in \(V_{\mathrm{bd}}\) of approximately 1--1.5~V is observed. In both FBK and HPK devices, \(V_{\mathrm{bd}}\) exhibits the expected temperature dependence, consistent with previous studies~\cite{currasrivera2025,COLLAZUOL2011389}, although a slight change in slope is seen at the highest fluence. For clarity, the small spread in \(V_{\mathrm{bd}}\) measured at 300~K between different modules of the same type was corrected by subtracting the difference of each module's \(V_{\mathrm{bd}}\) from the average \(V_{\mathrm{bd}}\) of that module type at 300~K, so that all curves start from a common reference and can be directly compared between devices and fluences.

\subsection{DCR as a function of temperature}
\label{sec:DCR_T}

The DCR for all types of SiPM was measured as a function of overvoltage and temperature across the full range of irradiated fluences. The measurements at each temperature were derived from the I--V characteristics, where the DCR was extracted from the ratio between dark current ($I_\mathrm{dark}$) and gain. The gain for all SiPMs at room temperature is known and its variation with temperature is small, as demonstrated in~\cite{currasrivera2025}. Although the reference reports that the gain changes by up to 20\% at low temperatures, this effect is small in this context. The observed variations in DCR span several orders of magnitude over the full temperature range. Consequently, assuming a constant gain does not affect the overall interpretation of the results. In addition, it is assumed that the gain does not vary with the irradiation fluence, a hypothesis confirmed by laboratory measurements with an estimated uncertainty of 10\%. To obtain a reliable average DCR for each module, multiple channels were measured within each array, and average values are reported.

\begin{figure}[t]
\centering
\includegraphics[width=0.80\columnwidth]{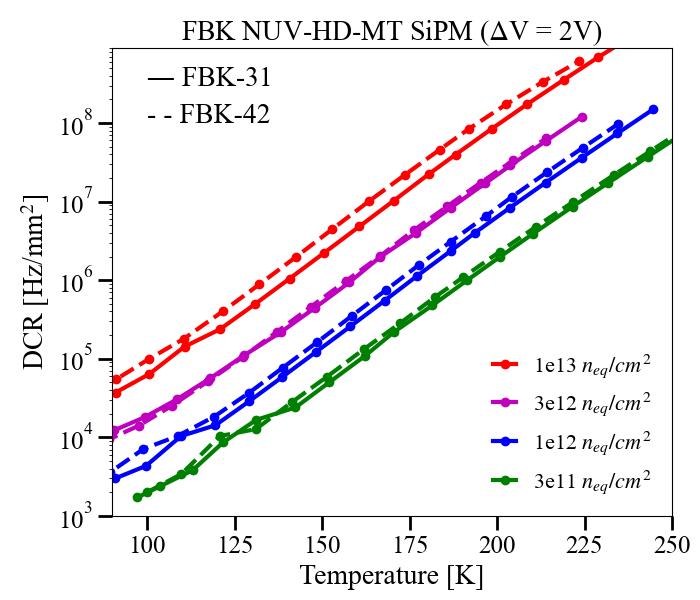}
\caption{DCR as a function of temperature for FBK SiPMs at an overvoltage of 2V. The curves represent fluences of \SI{3e11}, \SI{1e12}, \SI{3e12}, and \SI{1e13}{\neqcm}. Dashed lines represent the \SI{42}{\micro m} pixel size, while continuous lines represent the \SI{31}{\micro m} pixel size.}
\label{fig:DCR_FBK_2V}
\end{figure}

\begin{figure}[t]
\centering
\includegraphics[width=0.80\columnwidth]{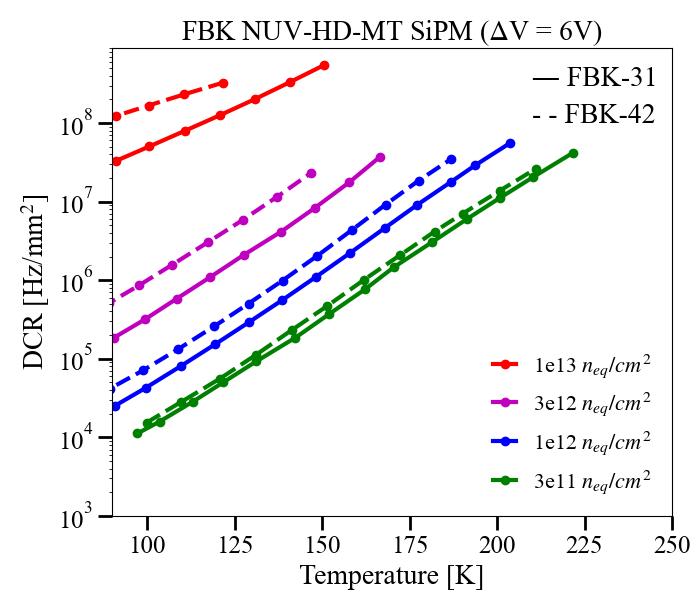}
\caption{DCR as a function of temperature for FBK SiPMs at an overvoltage of 6V. The curves represent fluences of \SI{3e11}, \SI{1e12}, \SI{3e12}, and \SI{1e13}{\neqcm}. Dashed lines represent the \SI{42}{\micro m} pixel size, while continuous lines represent the \SI{31}{\micro m} pixel size.}
\label{fig:DCR_FBK_6V}
\end{figure}

\begin{figure}[t]
\centering
\includegraphics[width=0.80\columnwidth]{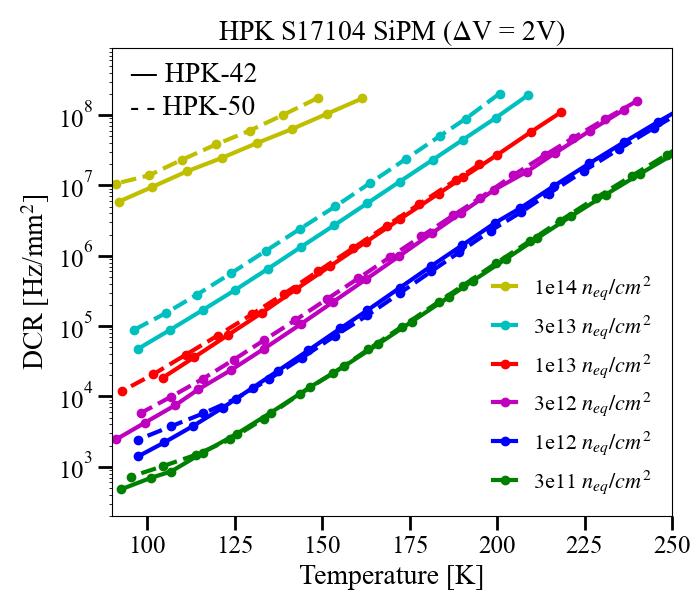}
\caption{DCR as a function of temperature for HPK SiPMs at an overvoltage of 2V. The curves represent fluences of \SI{3e11}, \SI{1e12}, \SI{3e12}, \SI{1e13}, \SI{3e13}, and \SI{1e14}{\neqcm}. Dashed lines represent the \SI{50}{\micro m} pixel size, while continuous lines represent the \SI{42}{\micro m} pixel size.}
\label{fig:DCR_HPK_2V}
\end{figure}

\begin{figure}[t]
\centering
\includegraphics[width=0.80\columnwidth]{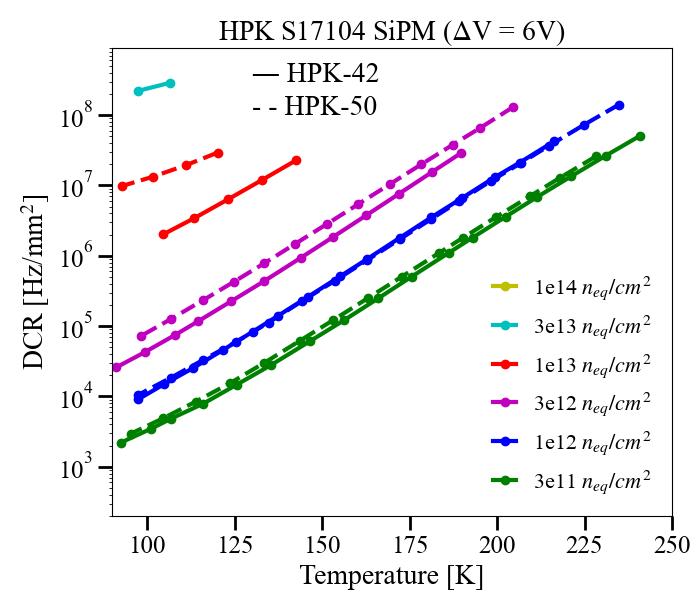}
\caption{DCR as a function of temperature for HPK SiPMs at an overvoltage of 6V. The curves represent fluences of \SI{3e11}, \SI{1e12}, \SI{3e12}, \SI{1e13}, \SI{3e13}, and \SI{1e14}{\neqcm}. Dashed lines represent the \SI{50}{\micro m} pixel size, while continuous lines represent the \SI{42}{\micro m} pixel size.}
\label{fig:DCR_HPK_6V}
\end{figure}

The DCR as a function of temperature for different overvoltages is shown in Figs.~\ref{fig:DCR_FBK_2V}--\ref{fig:DCR_HPK_6V}. For fluences up to \SI{1e12}{\neqcm}, the DCR decreases with cooling, following the expected dependence of Shockley--Read-Hall (SRH) trap-assisted generation on temperature~\cite{Sze2006,Piemonte2016}, with a halving factor of $K_{1/2} = 10.1~\mathrm{K}$. This behaviour, which corresponds to a thermally activated process in the depletion region, is consistent across all types of SiPMs for overvoltages below 6~V.
The results shown in Figs.~\ref{fig:DCR_FBK_2V}--\ref{fig:DCR_HPK_6V} are based on temperature scans where data points are recorded at a temperature interval of 10~K.

Two distinct operating regimes can be identified. At low overvoltage (for example, 2~V), the DCR increases proportionally with fluence for all SiPM modules studied, consistent with the Non-Ionising Energy Loss (NIEL) scaling hypothesis (proportional increase of DCR with fluence) of bulk damage~\cite{Vasilescu1997}. The only exception occurs for HPK SiPMs at the highest fluence (\SI{1e14}{\neqcm}), where the proportionality breaks down and the DCR increases faster than the expected rate. 

At high overvoltage (for example, 6~V), the proportional scaling of the DCR with fluence is maintained only up to \SI{3e12}{\neqcm}. Beyond this fluence, a significant increase in DCR beyond proportionality to fluence is observed. This behaviour is observed for both FBK and HPK SiPM arrays; depends on the applied overvoltage and is pixel size dependent. 

\begin{figure}[t]
\centering
\includegraphics[width=0.80\columnwidth]{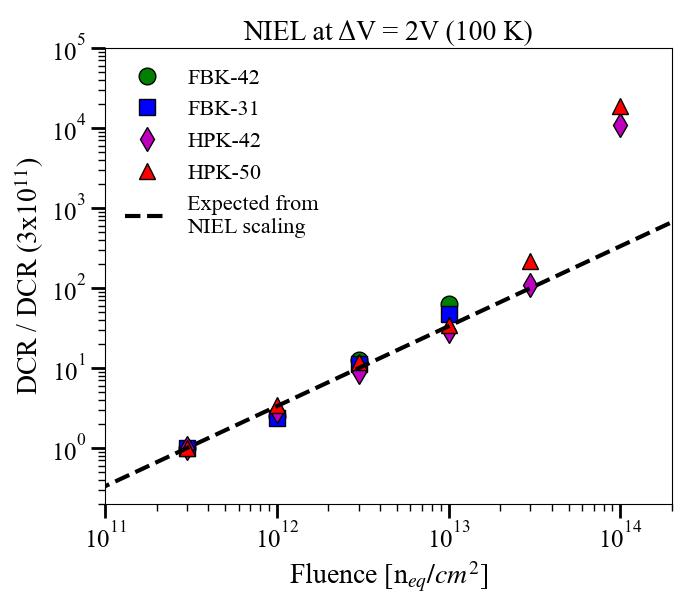}
\caption{The DCR at the lowest fluence \SI{3e11}{\neqcm} serves as the normalisation factor. DCR increases as a function of irradiation fluence for all SiPM array types at an overvoltage of 2~V. The proportional increase in DCR with fluence is observed up to approximately \SI{3e13}{\neqcm}, consistent with NIEL scaling.}
\label{fig:NIEL_2V}
\end{figure}

\begin{figure}[t]
\centering
\includegraphics[width=0.80\columnwidth]{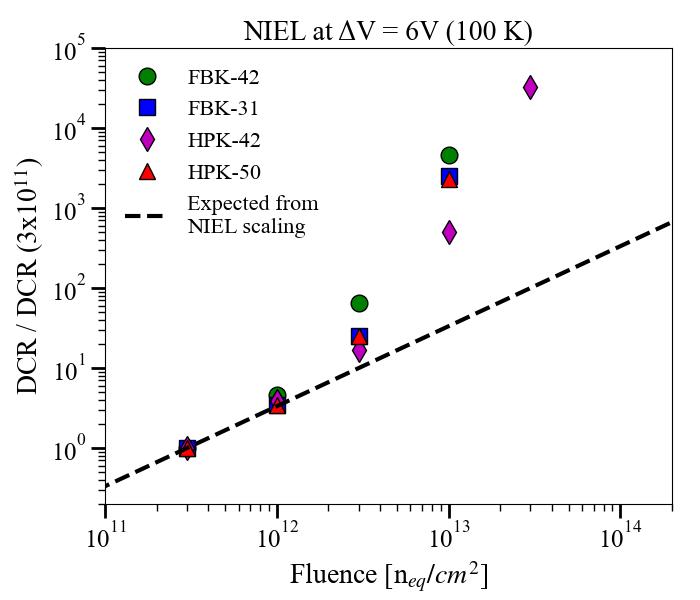}
\caption{The DCR at the lowest fluence \SI{3e11}{\neqcm} serves as the normalisation factor. DCR increases as a function of irradiation fluence for all SiPM array types at an overvoltage of 6~V. NIEL scaling holds only up to \SI{3e12}{\neqcm}; above this fluence, the DCR increases faster. The increase depends further on overvoltage and the pixel size.}
\label{fig:NIEL_6V}
\end{figure}

To evaluate the increase in DCR with respect to neutron fluence at a given overvoltage, the DCR ratio is plotted in Figs.~\ref{fig:NIEL_2V} and \ref{fig:NIEL_6V}. Here, the DCR at the lowest fluence \SI{3e11}{\neqcm} serves as a normalisation factor. 

\subsection{DCR as a Function of Overvoltage at 100~K}
\label{sec:DCR_overvoltage_100K}
In Section~\ref{sec:DCR_T}, it was observed that the NIEL scaling is limited to a range that depends on neutron fluence over a large temperature range. In this part, we present the DCR at a fixed temperature as a function of overvoltage to assess the field dependence of the DCR increase. 

\begin{figure}[t]
\centering
\includegraphics[width=0.90\columnwidth]{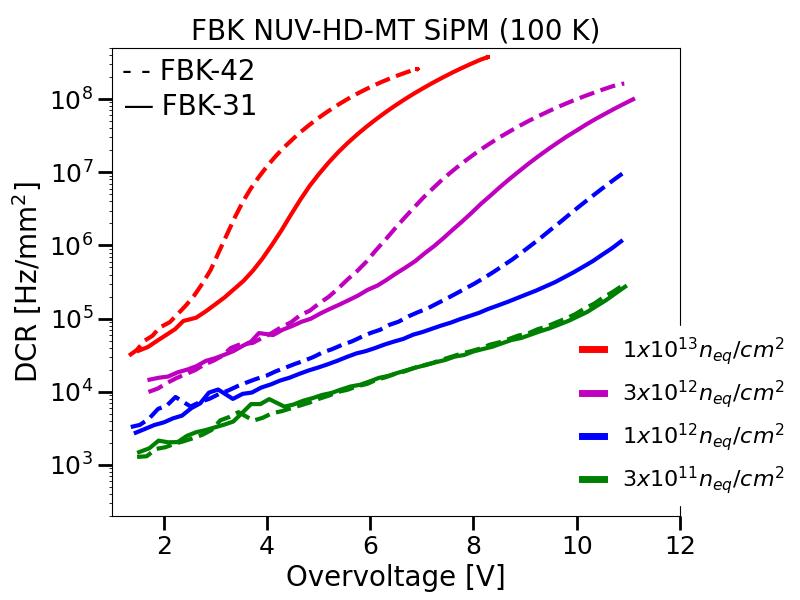}
\caption{DCR as a function of overvoltage for FBK at 100~K. For low fluences, the DCR increases linearly with overvoltage, while at higher fluences the DCR increase is much faster. Smaller pixels perform better.}
\label{fig:DCR_OV_FBK}
\end{figure}

\begin{figure}[t]
\centering
\includegraphics[width=0.90\columnwidth]{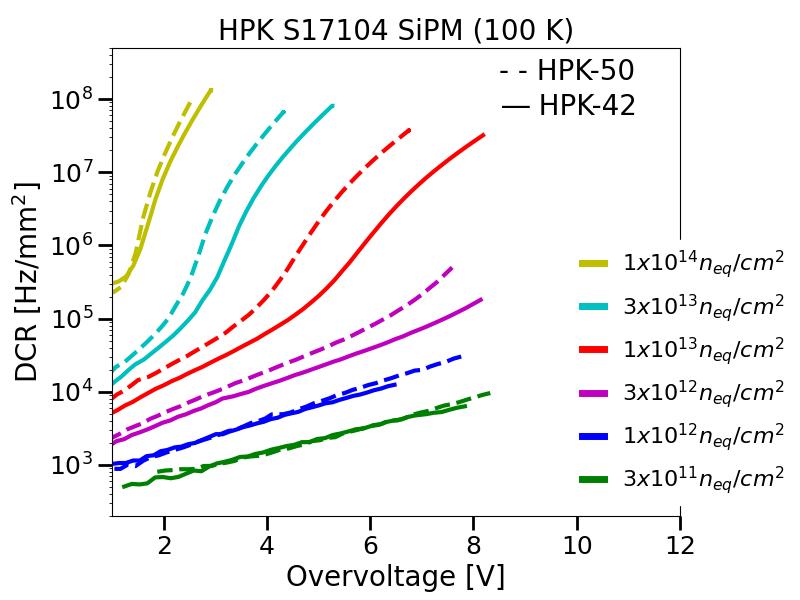}
\caption{DCR as a function of overvoltage for HPK at 100~K. For low fluences, the DCR increases linearly with overvoltage, while at higher fluences the DCR increase is much faster. Smaller pixels perform better.}
\label{fig:DCR_OV_HPK}
\end{figure}

Figs.~\ref{fig:DCR_OV_FBK}--\ref{fig:DCR_OV_HPK} show the DCR as a function of overvoltage. FBK and HPK devices are shown separately. At the lowest fluences and low overvoltage, the DCR increases approximately linearly with overvoltage on a logarithmic scale. However, at higher fluences and higher overvoltage, the linear behaviour changes into a much faster increase. The region in which the linear behaviour is observed depends on fluence, overvoltage, pixel size, and SiPM technology. 
In general, smaller pixels perform better. Upon comparison of the two technologies, FBK and HPK, HPK consistently shows a lower DCR than FBK for all fluences.

\subsection{DCR as a Function of Gain at 100~K}
\label{sec:DCR_gain_100K}

\begin{figure}[t]
\centering
\includegraphics[width=0.90\columnwidth]{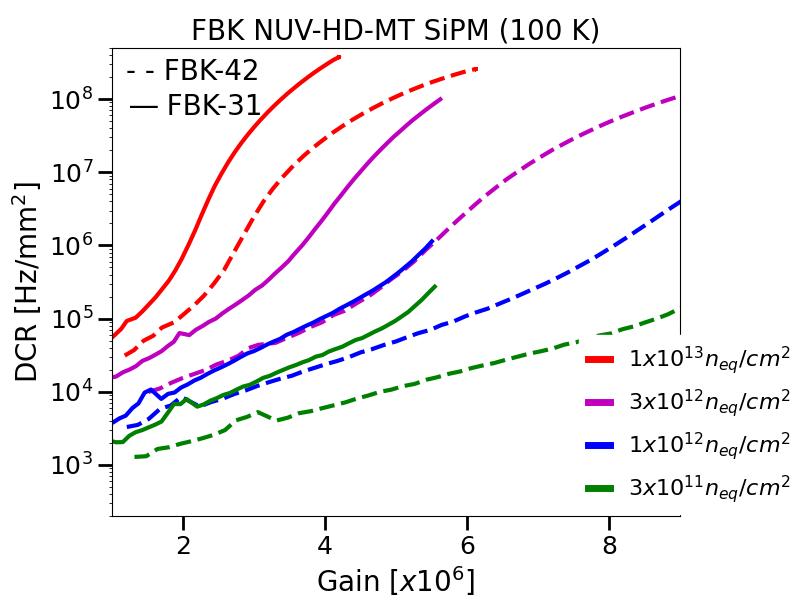}
\caption{DCR as a function of gain for FBK at 100~K. At low fluences, the DCR scales proportional with gain, while at higher fluences DCR increases much faster. Larger pixels perform better.}
\label{fig:DCR_Gain_FBK}
\end{figure}

\begin{figure}[t]
\centering
\includegraphics[width=0.90\columnwidth]{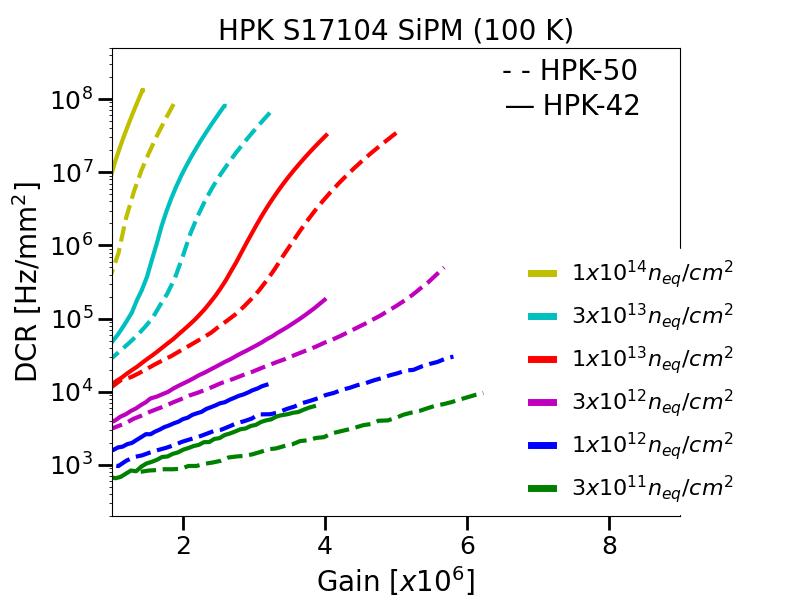}
\caption{DCR as a function of gain for HPK at 100~K. At low fluences, the DCR scales proportional with gain, while at higher fluences DCR increases much faster. Larger pixels perform better.}
\label{fig:DCR_Gain_HPK}
\end{figure}

Another key parameter of interest is the gain, and a similar analysis was performed. Figure~\ref{fig:DCR_Gain_FBK} and Figure~\ref{fig:DCR_Gain_HPK} present the DCR as a function of gain for the same SiPM arrays, again showing FBK devices separately from HPK devices. For low fluences, the DCR, in the logarithmic scale, scales linearly with gain, but at higher fluences this proportionality breaks down, and the DCR begins to rise exponentially beyond a certain gain threshold. In this case, SiPMs with smaller pixel sizes tend to reach higher DCR values than those with larger pixels. Consistent with the overvoltage results, HPK SiPMs display lower DCR than FBK devices for a given gain.

\subsection{DCR Activation Energy}
\label{sec:DCR_Ea}

\begin{figure}[t]
\centering
\includegraphics[width=0.90\columnwidth]{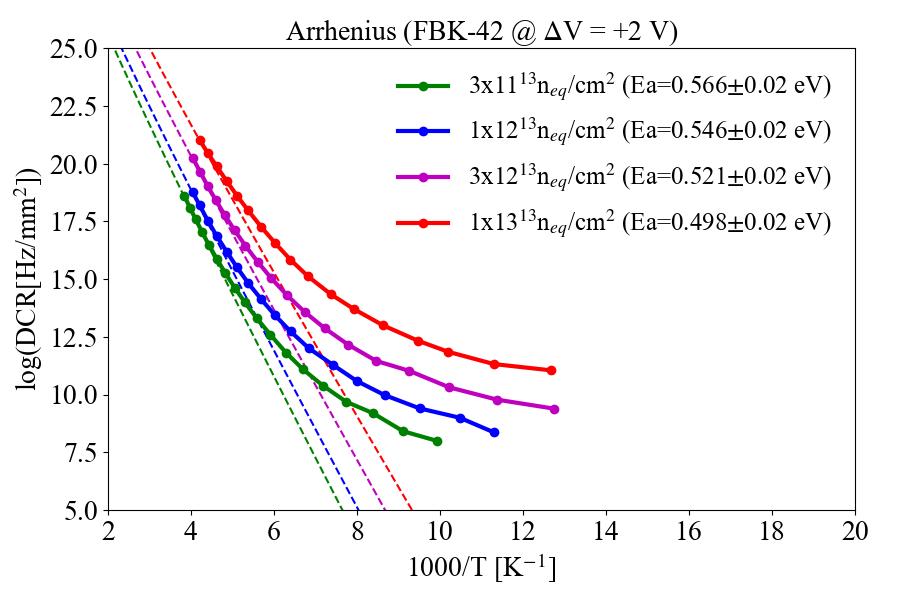}
\caption{Arrhenius plot of the DCR for the FBK-42 SiPM module, showing the extracted activation energies for different irradiation fluences. Measurements start at 250~K, so the diffusion component is not visible; only the SRH and tunnelling contributions are observed. The decrease in \(E_a\) with increasing fluence indicates enhanced radiation-induced defect generation.}
\label{fig:Ea_fluence}
\end{figure}

Although the DCR is often plotted as a function of temperature for clarity, its physical dependence follows an exponential trend with \(1/T\), making the Arrhenius representation more appropriate. The DCR can be expressed as
\[
\mathrm{DCR}(T) \propto \exp\left(-\frac{E_a}{kT}\right),
\]
where \(E_a\) is the activation energy associated with the dominant DCR generation process. The diffusion, Shockley-Read-Hall recombination (SRH), and tunnelling mechanisms exhibit progressively weaker temperature dependences, with typical activation energies of \(E_g\), \(E_g/2\), and less than \(E_g/2\), respectively. This behaviour is illustrated in Fig.~\ref{fig:Ea_fluence}, where the Arrhenius plot is shown for one of the FBK SiPM modules studied. With the measurement starting at 250~K, the diffusion component is not visible, but only the SRH and tunnelling contributions are observed. 

The extracted activation energies indicate that, at lower fluences, \(E_a\) remains close to \(E_g/2\), suggesting limited radiation-induced defect generation and that SRH is the dominant mechanism contributing to DCR. With increasing fluence, \(E_a\) decreases, reflecting enhanced field-assisted generation and more severe radiation damage. Note that the \(E_a\) values extracted from this graph are only indicative values, as the temperature range for the fit is very restricted.

\subsection{Annealing Effects on the DCR}
\label{sec:annealing}
As discussed in Section~\ref{sec:sipm_arrays}, all irradiated SiPM arrays underwent an initial annealing step of two weeks at $30\,^{\circ}\mathrm{C}$ before any measurements. This procedure reflects the operational conditions expected during the experiment, where the detectors will be repeatedly exposed to similar temperatures for extended periods during end-of-year shutdowns, maintenance, or technical stops. Approximately two weeks at this temperature are required to achieve the full recovery associated with beneficial annealing, during which certain defects recombine, leading to a reduction of the DCR.

To investigate whether additional annealing could further reduce the measured DCR, selected FBK-31 arrays were subjected to high-temperature treatments. This study was carried out for all irradiated fluences except \SI{3e11}{\neqcm}, as the corresponding module was damaged during one of the cooling cycles. The DCR was measured exclusively at 100~K, directly measuring the dark pulse rate by setting the detection threshold to half the amplitude of the single photoelectron signal, rather than extracting it from the dark current. Both measurement methods yielded comparable results, supporting the hypothesis discussed in Section~\ref{sec:DCR_overvoltage_100K} that the gain does not change significantly with cooling or the irradiation fluences investigated in this work.

\begin{figure}[t]
\centering
\includegraphics[width=0.99\columnwidth]{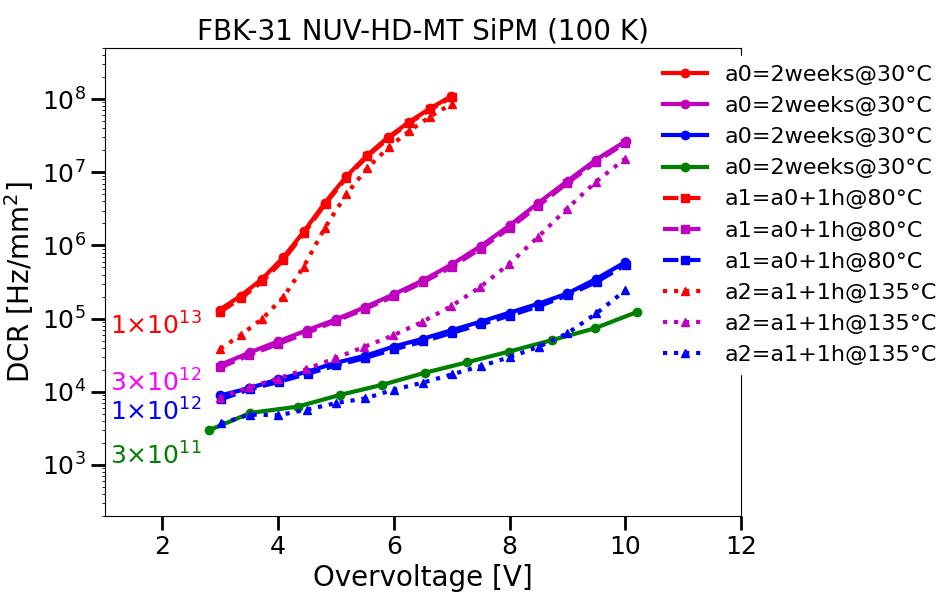}
\caption{Effect of annealing on the DCR of irradiated FBK-31 SiPM arrays measured at 100~K. The initial annealing at $30\,^{\circ}\mathrm{C}$ reduces the DCR through beneficial annealing. High-temperature annealing at $80\,^{\circ}\mathrm{C}$ shows negligible improvement, while annealing at $135\,^{\circ}\mathrm{C}$ significantly reduces the DCR at low overvoltages.}
\label{fig:annealing}
\end{figure}

The first annealing step was performed at $80\,^{\circ}\mathrm{C}$ for one hour, followed by DCR measurements at 100~K. Subsequently, the same modules underwent an additional annealing step at $135\,^{\circ}\mathrm{C}$ for one hour. 
The main findings are that only after the highest temperature annealing step ($135\,^{\circ}\mathrm{C}$), an additional significant reduction in DCR is observed. The initial annealing at $30\,^{\circ}\mathrm{C}$ reduced the observed DCR to the level where the intermediate annealing temperature step ($80\,^{\circ}\mathrm{C}$) does not have a visible effect.  
The highest temperature annealing step shows its full effect only at low overvoltages. At the highest overvoltage points, again, almost no effect is observed.
The complete set of results is presented in Fig~\ref{fig:annealing}.

\section{Discussion}
\label{sec:discussion}
The results presented in this work demonstrate that cryogenic cooling of SiPMs in a radiation environment can mitigate radiation-induced defects. Detectors cooled to 100~K perform as well as devices operated at room temperature, even after a fluence of \SI{3e13}{\neqcm}. However, DCR is strongly dependent on voltage, pixel size, and the technology used. The overvoltage and gain dependence of DCR generation limit the possible operating range of devices beyond \SI{3e12}{\neqcm}. 

For a neutron fluence below \SI{3e13}{\neqcm}, only minor differences between irradiated and non-irradiated devices are observed in terms of the temperature dependence of the breakdown voltage. With the highest fluence (\SI{1e14}{\neqcm}) differences were observed at the level of 1--1.5~V.
This observation suggests that neutron-induced defects primarily affect generation--recombination processes rather than the space-charge distribution in the avalanche region. The increase in $V_{\mathrm{bd}}$ observed at the highest fluence likely reflects radiation-induced modifications of the effective doping profile in the gain layer, leading to subtle adjustments of the local electric field. These changes may also influence other performance parameters that were not addressed in this study, such as PDE.

Cooling the detectors from room temperature to 100~K reduces the DCR by six orders of magnitude, allowing single photon detection even after irradiation levels as high as \SI{3e12}{\neqcm}, corresponding to those estimated for LHCb Upgrade~2. 
For fluences up to approximately \SI{3e12}{\neqcm}, the DCR follows the expected NIEL scaling, confirming that bulk defects dominate noise generation, i.e., the DCR and neutron fluence are proportional. Above \SI{3e12}{\neqcm}, a significant deviation from proportionality is observed. The most notable are those at higher overvoltages, indicating the onset of additional current components likely related to tunnelling-assisted free charge carrier generation and field-enhanced defect emission. These additional effects were seen with no temperature dependence and confirm the former hypothesis of their origins. At this fluence and beyond, an increase in correlated noise is likely present, and the increase in DCR is associated with afterpulsing. The noise is no longer only thermally generated but is correlated with avalanches earlier in time.  This type of generation cannot be reduced by cooling and is proportional not only to the thermal generated avalanches, but also to those produced by the signal. More research is needed to clearly identify the origin of this excess DCR.

At the highest fluences, the increase in DCR with overvoltage indicates that trap-assisted tunnelling becomes the dominant dark-count mechanism. This effect is particularly pronounced in devices with larger pixel sizes, consistent with their higher electric fields in the active region and larger junction areas. A comparison between the two technologies shows that HPK SiPMs maintain significantly lower DCR than their FBK counterparts across all fluences and operating conditions. 
Note that this evaluation only considers DCR leaving out other aspects as PDE and correlated noise.
This behaviour reflects differences in microcell design, surface passivation, or the spatial distribution of defects within the avalanche layer. The higher $V_{\mathrm{bd}}$ observed in HPK SiPMs suggests a thicker p-n junction, resulting in a lower electric field at the same overvoltage compared to the FBK devices. This supports the interpretation that the more pronounced excess DCR in FBK SiPMs is strongly linked to electric-field effects.

The observed annealing behaviour further supports the interpretation that a fraction of the radiation-induced defects can be annealed through thermal treatment. Annealing at $30\,^{\circ}\mathrm{C}$ leads to a substantial reduction in DCR (3.3 for LHCb SciFi Upgrade~1), whereas only a much higher temperature ($135\,^{\circ}\mathrm{C}$) allows for additional annealing by a factor of $\approx 3$ at low overvoltage. This suggests that some deep-level defects responsible for generation--recombination processes can be thermally deactivated without adversely affecting the breakdown or quenching behaviour. In contrast, the regime where trap-assisted tunnelling is likely to dominate shows little improvement after annealing, indicating that the DCR mechanisms in this region are fundamentally different.

Note that such high-temperature annealing is only compatible with fully assembled detector modules if properly engineered for the high temperature. A careful evaluation of long-term stability and system integration is required. 
In this context, more detailed annealing studies are planned, including a larger number of modules and a systematic exploration of different temperatures and annealing times. The ultimate goal is to achieve comparable reductions in DCR through forward-bias annealing, which selectively heats only the active region of the SiPM modules, minimising the thermal exposure for the organic plastic scintillating fibres closely coupled to the photodetector.

\section{Summary and Conclusions}
\label{sec:summary}

A comprehensive study of the performance of irradiated SiPM arrays operated at cryogenic temperatures has been presented. The measurements cover a wide range of fluences, temperatures, and overvoltages, allowing a detailed investigation of radiation-induced degradation mechanisms and their possible mitigation by low-temperature operation and thermal annealing.

The breakdown voltage followed the expected decrease with decreasing temperature and is largely unaffected by neutron irradiation, except at the highest fluence of \SI{1e14}{\neqcm}, where it increases approximately 1 to 1.5 \,V. This indicates that irradiation below this fluence mainly affects generation-recombination centres rather than the electric field distribution in the avalanche region. In contrast, the dark count rate shows a strong dependence on both temperature and irradiation. Cooling to 100~K reduces the DCR by several orders of magnitude, making cryogenic operation a highly effective strategy for suppressing radiation-induced noise. For moderate fluences up to \SI{3e12}{\neqcm}, the DCR follows the expected NIEL scaling, while at higher fluences and overvoltages, deviations consistent with trap-assisted tunnelling and field-enhanced defect emission are observed.

Performance comparison for FBK or HPK devices shows that larger pixels perform better at the same gain, while smaller pixels perform better at the same overvoltage. When the two technologies are compared at the same gain or overvoltage, HPK devices show consistently lower DCR. This suggests that differences in microcell design, avalanche region thickness, and material processing are important factors. However, FBK devices feature a significantly higher geometrical fill factor than similar pixel size devices from HPK. This leads to a higher PDE for the Scifi detector under similar operation conditions for FBK.

Annealing studies indicate that a portion of radiation-induced damage can be mitigated through thermal treatments. Beneficial annealing at $30\,^{\circ}\mathrm{C}$ results in a significant reduction in DCR, while high-temperature annealing at $135\,^{\circ}\mathrm{C}$ produces a further substantial improvement, limited to the low-overvoltage regime. In contrast, the component of DCR that may be associated with trap-assisted tunnelling appears to be largely unaffected, suggesting the presence of distinct defect mechanisms. 

In summary, the results confirm that cryogenic operation of SiPMs at a fluence of up to \SI{3e12}{\neqcm} can be performed in the single photon detection regime at moderate overvoltage. DCR can be as low as room temperature for non-irradiated devices. Cooling extends the use of these devices for the radiation environment of LHCb Upgrade~2. 

Future work will focus on the detailed characterisation of photon detection efficiency, timing, and correlated noise performance at cryogenic cooling with similar fluence to further evaluate the operational limits of these devices.

\section*{Acknowledgment}

This project has received funding from the European Union’s Horizon Europe Research and Innovation programme under Grant Agreement No 101057511 (EURO-LABS).

\bibliographystyle{ieee}

\end{document}